\documentclass[12pt,letterpaper,hyper]{JHEP3}
\usepackage[utf8]{inputenc}
\usepackage{epic,eepic}
\usepackage{amsmath,epsfig}
\usepackage{amssymb,amsfonts}
\usepackage{graphicx}

\title{The unconditional RG flow of the relativistic holographic fluid}

\preprint{}
\author{Stanislav Kuperstein and Ayan Mukhopadhyay\\

\it {Laboratoire de Physique Th\'eorique et Hautes
Energies (LPTHE)\\
\it{Universit\'e Pierre et Marie Curie -- Paris 6; CNRS UMR
7589}\\
\it{Tour 13-14, 4$^{\grave{e}me}$ \'etage, Boite 126, 4 Place
Jussieu}, 
\it {75252 Paris Cedex 05, France}}\\

\texttt{e-mails}: \textsf{skuperst, ayan@lpthe.jussieu.fr}\\
}

\abstract{
We study asymptotically slowly varying perturbations
of the AdS black brane in Einstein's gravity with a negative cosmological
constant. We allow both the induced metric and the Brown-York stress tensor at a given radial cut-off slice to fluctuate. These fluctuations, which determine the radial evolution of the metric, are parametrized in terms of boundary data. We observe that the renormalized energy-momentum tensor at any radial slice takes the standard hydrodynamic form which is relativistically covariant with respect to the induced metric. The RG flow of the fluid takes the form of field redefinitions of the boundary hydrodynamic variables. To show this, up to first order in the derivative expansion, we only need to investigate the radial flow of the boundary data and do not need to impose constraints on them. Imposing the constraints gives unforced nonlinear hydrodynamic equations at any radial slice. Along the way we make a careful study of the choice of counter-terms and hypersurfaces involved in defining the holographic RG flow, while at the same time we do not explicitly set any boundary condition either at the cut-off  or at the horizon. We find that $\eta/s$ does not change along the RG flow, equaling $1/(4\pi)$ when the future horizon is regular. We also analyze the flow of the speed of sound and find that it diverges at the horizon.
}
\keywords{Holography, Hydrodynamics, Renormalization Group, Black Holes}

\newcommand{\cut}{\mathbf{{\scriptstyle{c}}}}
\newcommand{\bd}{\mathbf{{\scriptstyle{b}}}}

\def\bea{\begin{eqnarray}}
\def\eea{\end{eqnarray}}
\def\be{\begin{equation}}
\def\ee{\end{equation}}
\def\ba{\begin{align}}
\def\ea{\end{align}}
\def\bse{\begin{subequations}}
\def\ese{\end{subequations}}

\newcommand{\bem}{\begin{pmatrix}}
\newcommand{\eem}{\end{pmatrix}}

\begin{document}

\section{Introduction}

Since the early days of AdS/CFT \cite{Maldacena, Polyakov, Witten}, it has been intuitively clear, 
that the radial direction which takes us away from the boundary of the space-times dual to the CFT states, 
encodes the CFT's renormalization group (RG) flow. However, a clear and general formulation of how the 
evolution of the boundary variables in the radial direction as given by Einstein's equations, realizes the 
field-theoretic renormalization group flow, still eludes us.

Instead of trying to gain a full understanding of the holographic renormalization group flow, here we will
try to test the correspondence in a regime where the gravitational flow equations become more tractable. 

Such a simplification indeed occurs in the long-wavelength regime, 
namely by considering only low frequency fluctuations in the field
theory directions. The Einstein equations then can be solved order by order in the derivative expansion, 
and the results may be compared with the \emph{hydrodynamical} expansion in the dual
field theory. This framework is known as \emph{the fluid/gravity correspondence}.

One might argue, though, that it is a bit counter-intuitive to talk of RG flow in the hydrodynamic regime,
since it is already a low-energy approximation.
However, there is a natural description of a renormalization flow of a non-linear partial differential equation 
like the Navier-Stokes' equation. It is known as \emph{Barenblatt's renormalization group} \cite{Goldenfeld, Barenblatt}  
and it's basic definition is as follows. Suppose a field variable follows a non-linear partial differential equation. 
When the field is \emph{coarse-grained} (averaged) at a given spatial and temporal resolution, this averaged field 
follows a similar partial differential equation but with different parameters described by an appropriate 
renormalization group flow. Constructing this renormalization group flow is mathematically a difficult problem and 
can be easily implemented only for Hamiltonian systems \cite{Goldenfeld, Barenblatt}; nevertheless, numerical evidence 
suggests that it can be implemented for KdV-Burgers' and non-relativistic Navier-Stokes' equations as well \cite{Chorin}. 
This renormalization group flow gives key insights into turbulence.

In the context of \emph{relativistic} Navier-Stokes' equation, we are not aware of any attempt for renormalization 
maintaining manifest relativistic covariance. However, let us argue that in order to implement this we also 
need to evolve the space-time metric along the RG flow. Suppose, we average the four-velocity vector $u^\mu$ field 
over a space-time cell $\mathcal{V}_x$ around a space-time point $x$ as below:
\begin{equation}
\overline{u}^\mu (x) = \frac{1}{\vert \mathcal{V}_x \vert} \int \limits_{x^\prime \in \mathcal{V}_x} d^4 x' u^\mu (x') \, .
\end{equation}
The averaged field $\overline{u}^\mu (x)$ will \emph{not} satisfy the normalization 
condition $\overline{u}^\mu h_{\mu\nu} \overline{u}^\nu = -1$ with $h_{\mu\nu}$ being the original 
metric in which the fluid is living. In order to satisfy this we need to consistently define an averaged metric as well.
The holographic renormalization group flow achieves this beautifully as it naturally integrates the scale dependent 
hydrodynamic variables and the background $d$-dimensional metric together into the metric of a $d+1$ space-time. 
It will be interesting  to explore in the future if the holographic RG indeed gives a 
realization of Barenblatt renormalization group flow of the relativistic Navier-Stokes' equation. 
However, here we will not attempt to make a field theoretic interpretation of the holographic renormalization 
of the fluid beyond these observations.

In this paper we study a special class of geometries which are asymptotically both AdS and slowly varying, 
and tend to a boosted black brane solution far in the past or in the future. In the literature, such space-times 
have been studied by promoting the boost four-vector, the Hawking temperature and the flat metric on radial slices 
characterizing these black brane solutions, to arbitrary slowly varying functions of the boundary coordinates. 
Using the Dirichlet boundary condition at the boundary and regularity at the future horizon, one gets a systematic 
fluid/gravity correspondence 
\cite{Policastro1, Policastro2, Janik1, Janik2, Janik3, Baier, Bhattacharyya1, Natsuume, myself1}. 
Here we allow both the induced metric and the Brown-York stress tensor at a chosen radial slice to fluctuate 
and find how we can capture these fluctuations in terms of the boundary data. This implies that we are \emph{not} 
imposing \emph{any explicit boundary condition} on the radial flow of the boundary data. 

In accordance with expectations, the radius of a chosen radial slice is interpreted as the scale at which the 
observer probes the dual field theory. The product of this chosen radius $\rho_0$ and the temperature field 
$T(x)$, \emph{i.e.} $\rho_0 T(x)$, is a conformal invariant. The data at $\rho=\rho_0$, which completely 
determines the radial flow, can be encoded at the boundary in terms functions of this invariant. 
The sliding energy scale of the renormalization group flow becomes the variable \emph{cut-off} radius 
that throughout this paper will be denoted by $\rho^\cut$.  

We study the renormalized Brown-York stress tensor of these space-times at an arbitrary cut-off $\rho=\rho^\cut$ 
and find that, up to first order in the derivative expansion, we can always find field redefinitions of the 
hydrodynamic variables which brings the energy-momentum tensor to the standard (Landau-Lifshitz) hydrodynamic form, 
without using the constraints imposed by Einstein's equations on the data at any radial slice. 
We can find these field redefinitions by solving the radial flow of the boundary data only. 
Thus the renormalization group flow of the fluid takes the form of field redefinitions of hydrodynamic variables. 
Also, imposing the constraints of Einstein's equations implies the usual unforced nonlinear hydrodynamic equations 
at any cut-off radial slice $\rho^\cut$. Further $\eta/s$ is the same at any $\rho^\cut$ 
irrespective of the \emph{explicit} choice of boundary conditions and equals $\mp 1/(4\pi )$ when 
the past/future horizon is regular\footnote{
We also point out that it is not completely irrelevant to talk about boundary conditions where fluctuations 
about equilibrium die in the past rather than in the future. In real time finite temperature field theory 
(i.e. in the Schwinger-Keldysh formalism) for instance, for local diagrammatic rules to work we need to take 
into account both the retarded and advanced propagator (along with the Keldysh propagator). 
The hydrodynamic limit of the advanced propagator has a pole in the upper-half plane corresponding to the 
flip in the sign of the viscosity. In real time gauge/gravity duality (which is still not a fully developed subject) 
one may expect similarly that the time-reversed boundary condition where the past horizons are regular 
should play a role.
}. 
We also find that the speed of sound diverges at the horizon 
just as in the membrane paradigm \cite{Damour1, Damour2, Znajek, Damour3, Price1, Price2, Damour4, Eling1, Eling2}. 

For the sake of explicitness, we will consider a five-dimensional space-time dual to a four-dimensional fluid 
throughout this paper, though our results can be readily generalized to any dimension.

In recent literature \cite{Strominger1, Strominger2, Compere, Strominger3, Cai} (see also \cite{Iqbal, Nabamita} 
for study of scale dependence of holographic response functions), the so-called ``Wilsonian approach to fluid/gravity duality" 
has already been investigated. However, some or all of the following features recur, which are assumptions of (\emph{i})
the Dirichlet boundary condition at the cut-off as in the traditional holographic renormalization, (\emph{ii}) 
regularity at the future horizon as in AdS/CFT at finite temperature, (\emph{iii}) specific algebraic properties of 
the space-time,\footnote{In \cite{Strominger3}, a Petrov type 1 condition has been imposed on the geometry to reduce 
the bulk degrees of freedom of gravity to the fluid variables at the cut-off.} (\emph{iv}) a non-relativistic limit, 
and (\emph{v}) the linearized approximation. We show here these above-mentioned assumptions are not necessary in 
getting the essentials of the fluid/gravity correspondence at any scale though they could be necessitated by other 
considerations. Moreover, $\eta/s$ is determined by the regularity at the past/future horizon alone at any scale 
irrespective of the explicit boundary condition at the cut-off. Our study is however restricted to asymptotically 
AdS space-times only.

The motivation to remove the restriction of the Dirichlet boundary condition comes from some other recent 
investigations on the holographic renormalization group flow in the literature. In particular, it has been argued 
that the counter-term action at the cut-off hypersurface in the traditional holographic  renormalization 
prescription \cite{Henningson, Balasubramanian, de Boer, Skenderis1, Bianchi, Skenderis2} should be replaced by a 
boundary action which is the result of integrating out the bulk degrees of freedom in the region between the boundary 
and the cut-off \cite{Polchinski, Faulkner} (see also \cite{Paulos}). In the saddle point approximation this boundary 
action should be obtained in conjunction with the bulk solution \cite{Faulkner} and should not be just derived from 
the latter.  The boundary action thus may affect the boundary condition at the cut-off. Making comparison with field 
theoretic renormalization scheme, a precise prescription can be given for a bulk free scalar field theory.

However, it is not clear to us how to adapt these approaches to gravity. Nevertheless, we will do a careful study of 
the choice of counter-terms and hypersurfaces involved in defining the holographic renormalization group flow. 
We will show that  in the context of the hydrodynamic energy-momentum tensor, by making some simple and reasonable 
assumptions, we can fix the choice of the counter-term and the hypersurface for the renormalization group flow 
of the holographic fluid at the first order in the derivative expansion. In our approach we parametrize arbitrary 
boundary conditions at the cut-off, but it could be that the counter-term implicitly necessitates a definite boundary 
condition at the cut-off. We will investigate this issue further in future research, however for present purposes, 
the counter-term suffices to define the renormalized energy-momentum tensor at the cut-off. 

In our approach, we also do not need the non-relativistic limit of the bulk solution. Our results are however, 
consistent with the non-relativistic limit, as such a limit of any relativistically covariant fluid gives 
incompressible Navier-Stokes equation \cite{Bhattacharyya3}. This has been obtained holographically in the literature 
at an arbitrary radial slice as well \cite{Bhattacharyya3, Strominger1}. 

We also drop the requirement of regularity at the horizon, in order to work with arbitrary boundary data. 
In the context of AdS/CFT, we should restore the regularity condition at the end of our calculation, so that the 
transport coefficients get uniquely specified. Our general approach may shed light on the possible holographic 
emergence of gravity itself \cite{Jacobson,  Padmanabhan1, Verlinde, Padmanabhan2}, by relating holographic 
renormalization group flow to absence of naked singularities in the bulk space-time.

Finally, it is to be noted that the choice of bulk coordinates plays an important 
role in defining the renormalization group flow. 
This is because the Brown-York stress tensor on a radial slice which gives us the scale dependent thermodynamic
quantities and hydrodynamic transport coefficients is not covariant under transformation of the five bulk coordinates. 
So, choice of a bulk coordinate system is indeed a choice of a holographic renormalization scheme. In this paper, 
we choose \emph{the Fefferman-Graham} (FG) coordinates. This coordinate system can be universally adapted for any 
asymptotically  AdS space-time. 
Also, the one-to-one correspondence between conformal transformations in the boundary and bulk 
diffeomorphisms is obtained in any such space-time by finding the bulk diffeomorphisms which preserve the 
FG form of the metric. Therefore the radial coordinate in the FG coordinate
system is a good candidate for the field theory energy scale.
In the specific cases relevant for the fluid/gravity correspondence, 
the bulk metric in the FG coordinates has a coordinate singularity at the horizon where the effective 
temperature diverges. 
This coordinate singularity signals an end of the renormalization group flow and may have an 
important physical significance. This has already been pointed out in \cite{Swingle}, where an attempt has been 
made to draw a parallel between holographic renormalization and entanglement renormalization by pointing out that 
both ends at a finite scale typically where the effective temperature diverges. In the latter case, it signifies 
that at this scale various parts of the system are disentangled from each other and we do not need to coarse grain 
any further. 

It should be stressed, that at any finite distance form the horizon the derivative expansion of the metric 
has a non-zero radius of convergence, and so we can trust the analysis in the FG coordinates, despite the coordinate
singularity on the horizon. We also check the 
regularity of our gravity solution by transforming it to the \emph{Eddington-Finkelstein} coordinates, 
which are regular at the horizon.

\subsection*{Organization of the paper}

In section 2, we outline the asymptotically slowly varying AdS space-times and how they encode perturbations 
of both the induced metric and the Brown-York stress tensor at the cut-off radial slice. In section 3, we 
investigate the choice of counter-terms and hypersurfaces involved in defining the holographic renormalization group 
flow. In section 4 we find the field redefinitions which bring the renormalized energy-momentum tensor to the standard 
hydrodynamic form and also find the transport coefficients at an arbitrary radial slice. Finally, we discuss some open 
questions which we will like to investigate in the future. In the appendix, we present the gravity solution
in the Eddington-Finkelstein coordinates.

\section{The asymptotically slowly varying AdS space-times}

We will be studying space-times which are: 
\begin{itemize}
\item asymptotically both $AdS_5$ and slowly varying, 
\item tend to the boosted black brane solution far in the past or in the future.
\end{itemize}
Such space-times in Einstein's theory of gravity can be characterized by boundary data, 
which are the boundary metric and the boundary energy-momentum tensor. 
More specifically, the boundary energy-momentum tensor is the Balasubramanian-Krauss stress tensor, 
which is the appropriately renormalized Brown-York stress tensor when the cut-off is taken to the boundary. 
We will denote the boundary metric as $g^{\bd}_{\mu\nu}$ and the boundary energy-momentum tensor as 
$t^{\bd}_{\mu\nu}$.
 
Instead of characterizing the space-times by boundary data, we will choose to characterize them by the data at a radial slice 
$\rho_0$. These data will be the induced metric and the Brown-York stress tensor at this radial slice. As \emph{we will not be specifying
boundary conditions anywhere}, we have to parametrize these data in terms of arbitrary functions which will depend on $\rho_0$.
This parametrization has to be consistent with the hydrodynamic limit, i.e. with the derivative expansion. We will see later in this
section how these can be achieved by promoting the coefficients of the non-normalizable and normalizable modes of the metric fluctuations 
to arbitrary functions of the dimensionless variable $\rho_0 T(x)$.

For the boosted black brane solution $g^\bd_{\mu\nu}=\eta^{\bd}_{\mu\nu}$ and the $5d$ metric can be 
characterized by four parameters: the Hawking temperature $T_{\rm H}$ and the three independent components 
of the unit constant boost four-vector $u^\mu$, which satisfies $u^\mu \eta_{\mu\nu} u^\nu =-1$. The stress 
tensor is:
\be
\label{Tflat}
t^\bd_{\mu\nu} = \mathcal{K} \frac{ \left(\pi T_{\rm H} \right)^4}{4}
  \left( \eta_{\mu\nu} + 4 u_\mu u_\nu \right)\, , \quad \text{where} \quad 
 \mathcal{K} \equiv \frac{l^3}{4\pi G_{\rm N}}  
\ee
and $l$ is the radius of $AdS_5$.

In the asymptotically slowly varying space-times discussed above, the parameters 
$\eta_{\mu\nu}$, $u^\mu$ and $T_{\rm H}$ can be promoted to arbitrary slowly varying functions of the boundary 
coordinates $x$ \cite{Bhattacharyya1, myself1, Bhattacharyya4}. 
First, we consider a four-dimensional metric $h_{\mu\nu}(x)$ such that it reduces to $\eta_{\mu\nu}$ 
far in the past or in the future. This metric is also slowly varying such that its curvature scale is small 
compared to the Hawking temperature of the initial or final black brane configuration of our space-times. 
Second, the boost four-vector $u^\mu$ is promoted to  $u^\mu(x)$, an arbitrary slowly varying function 
of the boundary coordinates, which is normalized with respect to the metric $h_{\mu\nu}(x)$:
\begin{equation}\label{fd1} 
u^\mu (x) h_{\mu\nu}(x) u^\nu (x) = -1.
\end{equation}
As $h_{\mu\nu}(x)$ reduces to $\eta_{\mu\nu}$ in the far past or the far future, $u^\mu (x)$ can be made to 
coincide with the constant black brane boost vector $u^\mu$ as well. Also, as we promote $u^\mu$ to a function 
of  boundary coordinates, without loss of generality, we can use a field definition of $u^\mu$ such that it is 
the local velocity of energy transport. This can be done by going to the Landau-Lifshitz frame \emph{locally}:
\begin{equation}\label{fd2}
t^\bd_{\mu\nu}(x) u^\nu(x) = - e(x) u_\mu (x) \, ,
\end{equation}
where $e(x)$ is the local energy density and the index in $u^\mu(x)$ is lowered with the metric $h_{\mu\nu}(x)$. 
Now the boundary energy-momentum tensor (\ref{Tflat}) is given instead by:
\be
\label{Tcurved}
t^\bd_{\mu\nu} (x) = \mathcal{K} \frac{ \left( \pi T(x) \right)^4}{4} \big( h_{\mu\nu} (x) + 4 u_\mu(x) u_\nu(x) \big) \, ,
\ee
and the $x$-dependent temperature $T(x)$ can be defined by:
\begin{equation}\label{fd3}
e(x) =\frac{3}{4}\mathcal{K}\Big(\pi T(x)\Big)^4 \, ,
\end{equation}
so that $T(x)$ coincides with the Hawking temperature of the black brane in the far past or in the far future. 

Notice that the background with the non-constant $h_{\mu \nu}(x)$, $u^{\mu}(x)$ and $T(x)$ will solve the 
$5d$ Einstein equations only to the \emph{zeroth} order in the derivative expansion. 
Already at the first order in this expansion
both the boundary metric and energy momentum tensor will receive corrections built from the derivatives
of $h_{\mu \nu}(x)$, $u^{\mu}(x)$ and $T(x)$.  

Let us focus on  the first derivative correction to the bulk metric. 
In this paper we will consider only the most general \emph{tensor} perturbations 
(which will be proportional to the hydrodynamic shear tensor). Here the word ``tensor'' refers to
the transformation properties under the spatial rotation group.
In our upcoming paper \cite{Upcoming} we will study scalar and vector perturbations showing that 
the conclusions of this paper will be unaffected.

As with any other field in the AdS background, there will be two modes for the bulk metric perturbation, 
the normalizable and the non-normalizable one. Let us denote the \emph{dimensionless} 
coefficients of these modes by $c_{\rm n}$ and $c_{\rm nn}$ respectively. 

Let us stress again that throughout this paper these coefficients will be \emph{arbitrary},
as we \emph{don't want to specify boundary conditions anywhere}. For the sake of completeness, at the 
end of this section we will show how a choice of $c_{\rm n}$ and $c_{\rm nn}$ sets specific boundary conditions
at a given radial slice.

The non-normalizable mode of the bulk metric does not vanish on the AdS boundary and the boundary metric changes
for non-zero $c_{\rm nn}$ \cite{myself1}:
\be
\label{gb}
g^{\bd}_{\mu\nu}(x) = h_{\mu\nu}(x) + \frac{c_{\rm nn}}{\pi T(x)} \sigma_{\mu\nu}(x) + \mathcal{O}(\epsilon^2) \, , 
\ee
where:
\begin{equation}
\sigma_{\mu\nu}(x) \equiv P_\mu^{\phantom{\mu}\lambda}(x)P_\nu^{\phantom{\nu}\sigma}(x)
 \big(  \nabla_\lambda u_\sigma + \nabla_\sigma u_\lambda \big)(x) 
  - \frac{2}{3}P_{\mu\nu}(x) \left( \nabla^\alpha u_\alpha \right)(x) \, , 
\end{equation}
\begin{equation}
P_{\mu\nu}(x) \equiv h_{\mu\nu}(x) + u_\mu (x) u_\nu (x) \, ,
\end{equation}
and the $\nabla_\mu$ covariant derivatives are calculated from $h_{\mu\nu}(x)$. Also, all the indices are 
lowered/raised with this metric/it's inverse. 
Further, here and in all of the upcoming formulae $\epsilon$ is the derivative expansion parameter. It is the 
ratio of the typical length scale of variation of the boundary fields $\left(h_{\mu\nu}(x), T(x), u^\mu (x) \right)$ 
to the inverse of the late or early time temperature.

Notice that after considering the full metric perturbation, $h_{\mu\nu}$ will stand for the boundary metric only 
when the non-normalizable mode is switched off. 
In general it is an useful auxiliary variable (along with $u^\mu$ and $T$) to obtain the 
general hydrodynamic metric.

The boundary energy momentum tensor also receives corrections, 
but unlike for the metric case, the new term is proportional to $c_{\rm n}$:
\be
\label{tb}
t^{\bd}_{\mu\nu}(x) = \mathcal{K}\Bigg(\frac{\left(\pi T(x)\right)^4}{4}\Big( h_{\mu\nu}(x) + 4u_\mu (x) u_\nu (x) \Big)
 - c_{\rm n} \cdot \left(\pi T(x)\right)^3\sigma_{\mu\nu}(x) +  \mathcal{O}(\epsilon^2) \Bigg)  \, .
\ee
Here (similar to the previous equation) the $\left(\pi T(x)\right)^3$ factor in the last term ensures that
$c_{\rm n}$ is indeed  dimensionless.
It is easy to verify that the new $t^{\bd}_{\mu\nu}$ is both conserved and traceless
with respect to $g^{\bd}_{\mu\nu}$, namely:
\begin{equation}
\label{cb}
{g^\bd}^{\mu\nu}\ t^{\bd}_{\mu\nu} = 0 + \mathcal{O}(\epsilon^2)\, .
\end{equation}
In proving this one has to use that $\sigma_{\mu}^{\phantom{\nu} \mu}=0$ and $u^{\mu} \sigma_{\mu\nu}=0$.
Notice that the tracelessness condition implies that the boundary fluid is conformal.\footnote{Recall that 
the conformal anomaly will show up only at the fourth order in the derivative expansion.} 


We can now present the transport coefficients of the boundary fluid
as a function of the local temperature \emph{and} the parameters $(c_{\rm n}, c_{\rm nn})$. 
To achieve this goal, we have to 
re-write the boundary energy momentum tensor (\ref{tb}) in terms of the proper boundary parameters. First,
notice that the boundary metric is $g^\bd_{\mu\nu}(x)$ and not $h_{\mu\nu}(x)$. Second, we have to address 
the normalization of the $4$-vector $u_{\mu}(x)$, which was earlier defined as $u_{\mu} h^{\mu\nu} u_{\nu} = -1$, 
while for the correctly normalized
boundary velocity vector ${u^{\bd}}^{\mu}$ one needs:
\be
\label{ubNORM}
{u^\bd}^\mu {g^\bd}_{\mu\nu} {u^\bd}^\nu = -1 \, .
\ee
It appears, though, that if we set:
\be
\label{ubu}
{u^\bd}^\mu (x) = u^\mu (x) + \mathcal{O} (\epsilon^2) \, ,
\ee
then the normalization (\ref{ubNORM}) will still be satisfied up to the second order in the derivative expansion. 
The physical tensor  $\sigma^{\bd}_{\mu\nu}$ should also be defined as:
\begin{equation}
\sigma^{\bd}_{\mu\nu}(x) = {P^\bd}_\mu^\lambda (x) {P^\bd}_\nu^\sigma (x) 
     \Big(\nabla^{\bd}_\lambda u_\sigma+ \nabla^{\bd}_\sigma u_\lambda \Big)(x) 
      - \frac{2}{3}P^{\bd}_{\mu\nu}(x)\Big(\nabla^{\bd \alpha} u_\alpha\Big)(x) \, ,
\end{equation}
with:
\begin{equation}
P^{\bd}_{\mu\nu}(x) = u^\bd_\mu(x) u^\bd_\nu(x) + g^{\bd}_{\mu\nu}(x) \, .
\end{equation}
In the above expressions $\nabla^\bd$ is constructed from $g^{\bd}_{\mu\nu}(x)$ and all the indices are 
lowered/raised with this metric/it's inverse.
Similarly we obtain that:
\be
\label{sbs}
\sigma^{\bd}_{\mu\nu} = \sigma_{\mu\nu} + \mathcal{O}(\epsilon^2).
\ee
Before proceeding further, let us introduce two new parameters:
\be
c_1 = c_{\rm nn} \, \quad \textrm{and} \quad c_2 = c_{\rm n} + \frac{c_{\rm nn}}{4} \, .
\ee
With the field redefinitions \eqref{gb}, \eqref{ubu} and \eqref{sbs}  the boundary energy-momentum tensor \eqref{tb} takes the following conformally covariant hydrodynamic form with respect to the metric $g^{\bd}_{\mu\nu}$:
\be
\label{tbp}
t^{\bd}_{\mu\nu}(x) = \mathcal{K}  \Bigg( \frac{\left(\pi T^{\bd}(x)\right)^4}{4}
           \Big( g^{\bd}_{\mu\nu}(x)  + 4u^{\bd}_\mu (x) u^{\bd}_\nu (x) \Big) 
      -c_2 \left(\pi T^{\bd}(x)\right)^3\sigma^{\bd}_{\mu\nu}(x)
        + \mathcal{O}(\epsilon^2) \Bigg)  \, .
\ee
Here we set the boundary temperature $T^\bd(x)$ as:
\be
T^\bd(x) = T(x) \, .
\ee
This condition is equivalent to the requirement that the boundary energy density $e^\bd(x)$ will have the same dependence on the 
temperature as in (\ref{fd3}) 
(this also explains why we have not considered any $\partial u$ or $\partial h$ corrections in the coefficient of the first term in (\ref{tb})). It follows that:
\be
e^{\bd} = \frac{3}{4}\mathcal{K}\Big(\pi T^{\bd}(x)\Big)^4 \, , 
\quad p^{\bd} = \frac{1}{4}\mathcal{K}\Big(\pi T^{\bd}(x)\Big)^4 \,, 
\quad s^{\bd} = \mathcal{K} \pi \Big(\pi T^{\bd}(x)\Big)^3 \, ,
\ee
where the last two identities follow immediately from (\ref{tbp}) and the thermodynamic relation $sT=e+p$.
We also learn from (\ref{tbp}) that the shear viscosity is:
\be\label{etabd}
\eta^{\bd} = \mathcal{K} c_2 \left(\pi T^{\bd}(x)\right)^3 \, \quad \textrm{and} \quad
 \frac{\eta^{\bd}}{s^{\bd}} = \frac{c_2}{\pi}  \, .
\ee
We would like to emphasize that in order to arrive at (\ref{tbp}) we did not have to use the equations of motion.

To complete this section let us report the full five-dimensional solution that asymptotes to the boundary metric (\ref{gb}) and the boundary energy-momentum tensor (\ref{tb}).
 
As we do not \emph{yet} insist on the regularity condition at the future horizon or at the past horizon, and we focus on the renormalization group flow of the energy-momentum tensor, the Fefferman-Graham coordinates will be more suitable for our investigations. Further, it is easier to work with arbitrary boundary data in the Fefferman-Graham coordinates. Importantly, our $5d$ metric can be translated to the ingoing Eddington-Finkelstein coordinates order by order in the derivative expansion, where the change of coordinates is also systematically corrected order by order in the expansion \cite{myself1}. The details are in Appendix.

We denote the Fefferman-Graham radial coordinate as $\rho$ and as before the boundary coordinates collectively as $x$. Following \cite{myself1}, we can readily solve the radial flow in these coordinates, and we find that up to first order in the derivative expansion, the bulk metric takes the following form:
\begin{eqnarray}
\label{metric}
ds^2 &=& \frac{l^2}{\rho^2} \Big( d\rho^2 + g_{\mu\nu}(\rho,x)dx^\mu dx^\nu \Big), \\\nonumber
g_{\mu\nu}(\rho, x) &=& A(\rho,x) \ h_{\mu\nu}(x) + B(\rho,x) \  u_\mu (x) u_\nu (x) - \frac{C(\rho,x) }{\pi T(x)} \sigma_{\mu\nu}(x) 
  + \mathcal{O}(\epsilon^2).
\end{eqnarray}
By definition the functions $A$, $B$ and $C$ are dimensionless and are given by:
\begin{eqnarray}\label{ABC}
&& A = 1+ \frac{\left(\pi T(x)\rho\right)^4}{4} \, , \qquad
   B = \left(\pi T(x)\rho\right)^4   \left( 1+ \frac{\left(\pi T(x)\rho\right)^4}{4} \right)^{-1} \, , \\ \nonumber
&& C =  - \left( 1+ \frac{\left(\pi T(x)\rho\right)^4}{4} \right) 
        \left(c_1 + 2 c_2 \log \left( \frac{1 - \left(\pi T(x)\rho\right)^4/4}{1 + \left(\pi T(x)\rho\right)^4/4} \right) \right).
\end{eqnarray} 
It is easy to see that (\ref{metric}) indeed reduces to (\ref{gb}) for $\rho=0$, namely at the boundary. On the other hand, the
horizon is given by $(\rho \pi T(x) )^4 = 4$. Notice that the horizon is not a constant radial slice, but its radial location varies slowly due to slow variation of the temperature field $T(x)$.

The constraints of Einstein's equations are satisfied by (\ref{metric}) provided that:
\be
{\nabla^\bd}^\mu t^\bd_{\mu\nu} = 0 \, ,
\ee
where again the index is raised by ${g^\bd}^{\mu\nu}$. Besides, had we not imposed \eqref{cb} by construction of $t^\bd_{\mu \nu}$,
Einstein's equations would have enforced it instead.

Remarkably, by translating the metric \eqref{metric} to the outgoing/ingoing Eddington-Finkelstein coordinates as shown in Appendix, we conclude we have regularity at the past/future horizon only when:
\be
\label{hreg}
c_2 = \mp \frac{1}{4} \, ,
\ee
so that:
\be
\frac{\eta^{\bd}}{s^{\bd}} = \mp \frac{1}{4\pi} \, .
\ee
We see that it is the physical shear viscosity which alone determines the regularity at the past/future horizon.

Before closing this section let us make a crucial comment regarding the coefficients $c_1$ and $c_2$, 
or alternatively $c_{\rm n}$ and $c_{\rm nn}$. 
In general, both $c_1$ and $c_2$ can be $x$-dependent. This dependence, however, should appear only through 
the hydrodynamical variables $h_{\mu \nu}(x)$, $u^{\mu}(x)$ and $T(x)$. In our first derivative approximation
only $T(x)$ is actually an option. This is so because it takes at least one derivative to produce a scalar from 
$h_{\mu \nu}(x)$ and $u^{\mu}(x)$, and, at the same time, $c_{1}$ and $c_{2}$ already multiply 
one derivative terms. This seems to create a contradiction, since these coefficients are, by construction, 
dimensionless and as such cannot depend only on $T(x)$. The puzzle is resolved if we assume that the coefficients
depend also on a certain value of the radial coordinate $\rho_0$. This radial parameter $\rho_0$ is a choice of scale, 
where we specify the data (the induced metric and the Brown-York stress tensor) that determines the radial flow. The combination $\rho_0 T(x)$ is dimensionless 
(and as such preserves conformal invariance at the boundary) and thus, still solving the equations of motion, 
we might have:
\be
\label{cCutOFff}
c_1 = c_1 (\rho_0 T(x)) \quad \textrm{and} \quad c_2 = c_2 (\rho_0 T(x)) \, 
\ee
in (\ref{gb}), (\ref{tb}) and all other formulae in this section.

As we have already mentioned, regularity fixes only $c_2$ leaving $c_1$ completely unconstrained.
The easiest way to obtain $c_1$ as a function of $\rho_0 T(x)$ is to impose the Dirichlet boundary condition on the metric at $\rho=\rho_0$. In this context the Dirichlet boundary condition implies that the first order fluctuation of the induced metric at the radial slice $\rho = \rho_0$ vanishes. A simple inspection of \eqref{metric} shows that it leads to:
\be\label{cdirich}
c_1 \Big( \rho_0 T^{\bd}(x) \Big) = -c_2\Big( \rho_0 T^{\bd}(x) \Big) 
  \log \Bigg(\frac{1 - \left(\pi T^{\bd}(x)\rho_0 \right)^4/4}{1 + \left(\pi T^{\bd}(x)\rho_0 \right)^4/4}\Bigg),
\ee
and in conjunction with \eqref{hreg} this determines $c_1 \left( T^{\bd}(x)\rho_0 \right)$ completely. 

Our approach, however, will be essentially different.
We will show that for the fluid/gravity correspondence at any hypersurface, \emph{both} the Dirichlet boundary condition at the cut-off and the regularity at the past/future horizon should \emph{not} be explicitly imposed. In other words, the holographic renormalization group flow preserves the fluid/gravity correspondence for arbitrary $c_1$ and $c_2$. At any radial hypersurface, with appropriate field redefinitions, we will always get unforced hydrodynamic equations exactly as we have got here at the boundary. To arrive at this observation, it will suffice to determine only the stress tensor counter-term, without addressing the boundary conditions at all. We will see in the next section that with our approximation and with some simple additional assumptions this counter-term can be fixed unambiguously.

With this been said, there is still a possibility that in an implicit way our counter-term necessitates a \emph{definite} boundary condition at a radial hypersurface. We hope to address this issue in future research.

\section{Counter-terms and the renormalized energy-momentum tensor}

We recall the traditional prescription for \emph{holographic renormalization} \cite{Henningson, Balasubramanian, de Boer, Skenderis1, Bianchi, Skenderis2}. In this prescription, variables like the Brown-York stress tensor are evaluated first at a radial cut-off. This is morally dual to the field-theoretic procedure of regularization where we put an explicit \emph{cut-off} to tame ultraviolet divergences. However, these variables, evaluated at the cut-off radial slice, diverge at the boundary, and so just like in field theory we need \emph{counter-terms}.

In the traditional prescription the counter-terms that renormalize the Brown-York stress tensor are derived from a counter-term action living at the cut-off. This is in turn constructed by requiring that it cancels the divergences in the on-shell bulk action. However, we should also have a well defined variational principle. This is achieved by imposing \emph{the Dirichlet boundary condition} at the cut-off. This may require to introduce additional counter-terms like the Gibbons-Hawking counter-term which are not related to cancellation of divergences. 
In order to be consistent with the variational principle, the counter-term action required to cancel the UV divergences should be a functional of the values of bulk fields at the cut-off, their covariant tangential derivatives, but not of their radial derivatives.
Further, the traditional prescription also requires that the coefficients of all the terms appearing in this counter-term action should have the same values for all bulk solutions. 

As mentioned in Introduction, recent studies indicate that the coefficients of the counter-terms may \emph{depend} on the bulk solution \cite{Polchinski, Faulkner}. The boundary (counter-term) action is a result of integrating bulk degrees of freedom in the region between the boundary and the cut-off, and still should be a functional of the values of the bulk fields at the cut-off and their tangential derivatives. Nevertheless, it should be obtained in conjunction with the bulk solution in the saddle-point action and not just derived from the latter. Thus the boundary action affects the boundary condition at the cut-off. 

However, it is not clear to us how to implement such ideas for gravity perturbations in the bulk. At the same time, as we do not want to use the Dirichlet boundary condition at the cut-off, we also cannot use the traditional prescription of holographic renormalization. Another complication stems from the question that, as the metric gets perturbed, whether the correct choice of the cut-off hypersurface it still the one which is at a fixed radial location. 

Despite these difficulties, we can make a lot of progress by the following considerations without constructing an explicit boundary action:
\begin{itemize}
\item The counter-terms should reduce to the counter-terms of the traditional prescription at the boundary. This should be so because the ultraviolet divergences are typically independent of the state, for instance the ultraviolet divergences do not depend on the temperature in thermal field theory. The coefficients of the counter-terms at the boundary are thus the same for all bulk solutions. In other words we have to recover the traditional results at the boundary, for instance the Balasubramanian-Krauss energy-momentum tensor.

\item Away from the boundary the coefficients of the counter-terms in our space-times can depend on $\rho^\cut T(x)$ and $\rho^\cut \nabla^\cut$, with $\nabla^\cut$ being the covariant derivative constructed from the induced metric at the cut-off hypersurface. As in the holographic dual picture, we capture not the operator but its expectation value in a state, the coefficients of the counter-terms can have a dependence on the state, or equivalently the dual bulk solution. Also, $\rho^\cut T(x) \sim T(x)/\Lambda^\cut$ and $\rho^\cut \nabla^\cut \sim k/\Lambda^\cut$ if we make field-theoretic analogies.

\item We introduce all possible counter-terms which depend on the values of the fields at the cut-off and their covariant tangential derivatives, even if they are not required for canceling ultraviolet divergences. It is plausible that some of the coefficients may vanish at the boundary.
 
\item We require that the renormalized expectation value of the operator takes the same form at the cut-off hypersurface after appropriate field redefinitions. Furthermore, these field redefinitions should be obtained off-shell, namely without resorting to the equations of motion. This is typically true if there is an underlying path integral formalism for renormalization.

\item Imposing the energy-momentum conservation at the boundary should lead to the analogous conservation at the cut-off. 
\end{itemize}

Importantly, our discussion above is general and does not rely on the choice of a renormalization scheme. 
Instead we try to see what counter-terms one can add on the gravity side which conform with expectations 
on the field theory side. We admit that our discussion does not lead to explicit form of the counter-terms 
in the gauge theory.

It turns out that the above requirements specify the counter-terms to a large extent. Up to first order in the derivative expansion, the only counter-term we can add to the Brown-York stress tensor at the cut-off is proportional to:
\begin{equation}
f\Big(\rho^\cut T(x),\rho^\cut \nabla^\cut\Big)\gamma^\cut_{\mu\nu} \sim f\big(\rho^\cut T(x)\big)\gamma^\cut_{\mu\nu} \,  ,
\end{equation}
where we can drop the dependence on $\rho^\cut \nabla^\cut$ as the covariant derivative of the metric vanishes. Here 
$\gamma^\cut_{\mu\nu}$ is the induced metric at the cut-off $\rho=\rho^\cut$. The next counter-term that we can add should involve $Ric[\gamma]$ or $R[\gamma]$ and hence it contributes two derivative onward. We find this counter-term satisfies:
\begin{equation}
{\nabla^\cut}^\mu \Big( f \left( \rho^\cut T(x) \right) \gamma^\cut_{\mu\nu} \Big) = f^\prime \big( \rho^\cut T(x) \big)\partial_\nu T(x) \, ,
\end{equation}
where ${}^\prime$ denotes differentiation with respect to $\rho^\cut T(x)$. We find that as long as $f \left(\rho^\cut T(x)\right)$ is not constant
the new term violates the energy-momentum conservation.
To summarize, $f\left(\rho^\cut T(x)\right)$ should be a constant.\footnote{
Notice that the situation might change if we use other kinds of hypersurfaces like $\rho = a/T(x)$, where $a$ is an arbitrary numerical constant. On these hypersurfaces, we can introduce the counter-term $f(a)\gamma^\cut_{\mu\nu}$, where now the coefficient $f(a)$ could be a non-trivial function. As $a$ is a constant we get ${\nabla^\cut}^\mu \Big(f(a)\gamma^\cut_{\mu\nu}\Big)=0$, so we do not violate the Ward identity. We require that when $a=0$, the coefficient $f(a)$ is fixed by cancellation of the ultraviolet divergence in the Brown-York stress tensor. Unfortunately, we find that for such hypersurfaces, the renormalized energy-momentum tensor has derivatives of the temperature field, and no field redefinition can bring the hydrodynamic energy-momentum tensor to the standard form, unless we use the hydrodynamic equations of motion. Thus the renormalized energy-momentum tensor preserves its form on-shell but not off-shell as we change $a$.}

Notice that the physical metric at a given cut-off is not the induced metric itself. In the Fefferman-Graham coordinates it is:
\be\label{gc}
g^\cut_{\mu\nu}  = \frac{{\rho^\cut}^2}{l^2} \cdot \gamma^\cut_{\mu\nu} \, .
\ee
It coincides with the boundary metric $g^\bd_{\mu\nu}$ at $\rho=0$\footnote{
The multiplying factor $\rho^2/l^2$ above can be replaced by an arbitrary function of $\rho/l$ vanishing like $\rho^2/l^2$ at the boundary. This represents an overall choice of scale and does not affect any dimensionless physical quantity like the speed of sound or $\eta/s$ at any radial slice. }.
Clearly the covariant derivative $\nabla^\cut$ constructed from the induced metric $\gamma^\cut_{\mu\nu}$ is the same as that constructed from the physical metric $g^\cut_{\mu\nu}$.

The Brown-York stress tensor is given by:
\begin{equation}
T^{\rm Brown-York}_{\mu\nu} = - \frac{\mathcal{K}}{2 l^3} \left( K_{\mu \nu} - K \gamma_{\mu \nu} \right) \, .
\end{equation}
We emphasize that the Brown-York stress tensor is conserved with respect to the induced metric at any hypersurface irrespective of the boundary condition, as this is simply equivalent to the pullback of Einstein's equations to the cut-off. Therefore it acts like a bare energy-momentum tensor. In the Fefferman-Graham coordinates:
\be
K_{\mu \nu} = - \frac{l}{2} \rho \frac{\partial}{\partial\rho} \left( \frac{g_{\mu \nu}}{\rho^2} \right) \, .
\ee 
Up to first order in the derivative expansion, as discussed here, the counter-term for the energy-momentum tensor at a constant radial slices is:
\be
\label{tct1}
T^{\rm counter-term}_{\mu\nu} = - \frac{\mathcal{K}}{l^4}\Bigg(\frac{3}{2} \gamma_{\mu\nu}  + \mathcal{O}\left( \epsilon^2 \right)\Bigg) \, ,
\ee
where the coefficient is fixed by requiring the cancellation of the ultraviolet divergence of the Brown-York tensor at $\rho=0$. 

The physical renormalized energy-momentum tensor at a given cut-off is:
\begin{equation}\label{tren}
t_{\mu\nu} (\rho,x)=  \frac{l^2}{\rho^2} \left( T_{\mu \nu}^{\rm Brown-York} + T_{\mu \nu}^{\rm counter-term} \right).
\end{equation}
The extra multiplying factor of $l^2/\rho^2$ is determined by the multiplying factor $\rho^2/l^2$ used in defining the physical metric in terms of the induced metric in \eqref{gc} and is the same as in the traditional prescription. 

We find that with the bulk metrics \eqref{metric} under consideration, in the Fefferman-Graham coordinates the physical energy-momentum tensor up to first order in the derivative expansion at the cut-off radial slice is:
\bea\label{tc}
t^\cut_{\mu\nu} &=& \mathcal{K} \Big( \alpha \left( z^\cut \right) \big(\pi T^\bd(x)\big)^4 h_{\mu\nu}(x) 
                   + \beta \left( z^\cut \right) \big(\pi T^\bd(x) \big)^4 u_{\mu}(x) u_{\nu}(x)  \\\nonumber
                   && \qquad - \gamma\left( z^\cut \right) \big(\pi T^\bd(x) \big)^3 \sigma_{\mu\nu}(x) + \mathcal{O} \left( \epsilon^2 \right)\Big)  \,  ,
\eea
with:
\bea
\label{alphabetagamma}
\alpha(z) &=& \frac{A(z)^2}{A(z)-B(z)} \, \partial_z \left( \frac{B(z)}{A(z)} \right) -3  \partial_z A(z) \, , \\\nonumber
\beta(z) &=& \frac{A(z)^2}{A(z)-B(z)} \, \partial_z \left( \frac{B(z)}{A(z)} \right) -3  \frac{B(z)}{A(z)}\partial_z A(z) \, , \\\nonumber
\gamma(z) &=& \Big( A(z)-B(z) \Big) \, \partial_z \left( \frac{C(z)}{ A(z)-B(z) } \right) -3  \frac{C(z)}{A(z)}\partial_z A(z) \, ,
\eea
where $z$ and $z^\cut$ and defined by: 
\be
\label{z}
z = \left( \pi T^\bd (x) \rho \right)^4  
\quad \textrm{and} \quad
z^\cut = \left( \pi T^\bd (x) \rho^\cut \right)^4  \, .
\ee
However, we still need to find the physical hydrodynamic variables at the cut-off and demonstrate that the physical energy-momentum tensor takes the standard hydrodynamic form which is covariant with respect to the physical metric at the cut-off. We will do so in the next section.

\section{The RG flow of the fluid}

In the previous section we have found that at an arbitrary radial slice, the physical metric $g^\cut_{\mu\nu}(\rho,x)$ is \eqref{gc} and the physical energy-momentum tensor $t^{\cut}_{\mu\nu}(\rho,x)$ is \eqref{tc}.

Given that $t^{\cut}_{\mu\nu}$ takes the hydrodynamic form at the boundary, it is certainly not obvious that it will take that form at any radial slice. For this purpose we need to determine the appropriate hydrodynamic variables at the cut-off, ${u^\cut}^{\mu}(x)$ and $T^\cut(x)$.
 
Following the discussion in Section 2, we define the velocity field $u^{\cut}_{\mu\nu}$ with the following field definition (analogous to that at the boundary):
\be
\label{def1}
t^{\cut}_{\mu\nu}  {u^\cut}^\nu = - e^{\cut} g^\cut_{\mu \nu} {u^\cut}^\nu
\quad \textrm{and} \quad
{u^\cut}^{\mu} {g^\cut}_{\mu\nu} {u^\cut}^{\nu}  = -1.
\ee
This definition ensures that ${u^\cut}^\mu$ is the local velocity of energy transport and is normalized with respect to the physical metric ${g^\cut}_{\mu\nu}$ as at the boundary. 

The temperature field at the cut-off $T^{\cut}$ is red-shifted with respect to the temperature at the boundary by the factor $\sqrt{-g_{\mu\nu}u^\mu u^\nu}$, which is the square root of the local time-time component\footnote{Notice that ${u^\cut}^\mu \partial_{\mu}$
is the time-like Killing vector in the un-perturbed black brane background.} 
of the physical metric. Therefore:
\begin{equation}\label{def2}
T^{\cut} (\rho^\cut,x) = \frac{T^\bd(x)}{ \Big( A \left(  \rho^\cut T^\bd(x) \right)-B \left(  \rho^\cut T^\bd(x) \right) \Big)^{1/2} } 
     + \mathcal{O} \left( \epsilon^2 \right)\, .
\end{equation}
At the boundary ($\rho^\cut=0$) we have $A=1$ and $B=0$, so clearly $T^{\cut}$ coincides with $T^\bd$. On the other hand, on the horizon 
($\rho^\cut T^\bd = \sqrt{2}$) the red-shift factor diverges and
hence so $T^{\cut}$. Using the explicit form of $A$ and $B$ from \eqref{ABC} one can express $T^\bd$ in terms of $T^\cut$ and $\rho^\cut$.
Unfortunately, the final expression is not particularly illustrative, as it includes a root of a cubic polynomial, and
we will not report it here.

We have to find whether with these field definitions of ${u^\cut}^{\mu}$ and $T^{\cut}$, the physical energy-momentum tensor $t^{\cut}_{\mu\nu}$ at an arbitrary radial slice takes the conformally covariant hydrodynamic form with respect to the metric $g^\cut_{\mu\nu}$. 

Using $\sigma_{\mu\nu}u^\nu = 0$, as evident from the definition of $\sigma_{\mu\nu}$, we find that $u^\mu$ satisfies the first equation of \eqref{def1}, but does not satisfy the normalization condition. This implies that $u^\mu$ 
and ${u^\cut}^\mu$ are proportional to each other up to a normalization factor. We find:
\be
\label{ucu}
{u^{\cut}}^\mu (\rho^\cut,x) = \frac{{u^\bd}^\mu (x)}{ \Big( A \left(  \rho^\cut T^\bd(x) \right) 
                                                   - B \left(  \rho^\cut T^\bd(x) \right) \Big)^{1/2} } + \mathcal{O} \left( \epsilon^2 \right) \, ,
\ee 
where we used \eqref{ubu}. Non-surprisingly the time-like vector $u^\mu$ scales exactly as the temperature in \eqref{def2} and the field definition becomes singular at the horizon \footnote{This indicates that the appropriate four-velocity vector at the horizon is light-like as in the membrane paradigm \cite{Damour1, Damour2, Damour3, Eling1, Eling2}.}.

\textit{It should be observed that the field redefinitions that take us from the boundary fluid variables to the fluid variables at the cut-off are not conformal transformations, 
as the induced metric at the cut-off and the boundary metric are not related by any conformal transformation.}  We will also see that the hydrodynamic energy-momentum tensor at the 
cut-off is also not traceless with respect to the induced metric 
(up to first order in derivative expansion we can ignore considerations of the conformal anomaly, and claim that if the energy-momentum tensor is confomal, it should be traceless).

For the rest of the section we use the convention that whenever the superscript $\cut$ is shown explicitly, the indices are lowered/raised using the metric $g^\cut_{\mu\nu}$/it's inverse.
The appropriate definition of $\sigma^{\cut}_{\mu\nu}$ is given by:
\begin{equation}
\sigma^{\cut}_{\mu\nu}(x) = {P^\cut}_\mu^{\phantom{\mu}^\lambda} (x) {P^\cut}_\nu^{\phantom{\mu}^\sigma} (x)
     \left(\nabla^{\cut}_\lambda u^{\cut}_\sigma + \nabla^{\cut}_\sigma u^{\cut}_\lambda \right)(x) -       
     \frac{2}{3}P^{\cut}_{\mu\nu}(x)\left({\nabla^\cut}^\alpha u^{\cut}_\alpha\right)(x) \, , 
\end{equation}
with:
\begin{equation}
P^{\cut}_{\mu\nu}(x) = g^\cut_{\mu\nu}(x) + u^{\cut}_\mu(x) u^{\cut}_\nu(x) \, ,
\end{equation}
being the projection tensor on the hypersurface orthogonal to $u^\cut_{\mu}$ with respect to the metric $g^\cut_{\mu\nu}$ and $\nabla^{\cut}$ being the covariant derivative 
constructed from $g^\cut_{\mu\nu}$. 
Remarkably, using \eqref{ucu}, 
after some algebra, we find that:
\be
\label{scs}
\sigma_{\mu\nu}^\cut = \frac{A}{ \left( A-B \right)^{1/2}} \sigma_{\mu\nu} + \mathcal{O}\left( \epsilon^2 \right) \, .
\ee
The result above follows solely from the definitions of $\sigma_{\mu\nu}$ and $\sigma_{\mu\nu}^\cut$ and in proving it we did not have to resort to equations of motion.

It follows from \eqref{ucu} and \eqref{scs} that the physical energy-momentum tensor at the cut-off \eqref{tc} takes the standard covariant hydrodynamic form up to first order in the 
derivative expansion: 
\bea
\label{TmunuCutoff}
t^\cut_{\mu\nu} &=& \mathcal{K} \Big( \alpha^\cut \big(\rho^\cut T^\cut(x) \big) \big(\pi T^\cut(x) \big)^4 g^\cut_{\mu\nu}(x) + 
                                      \beta^\cut \big( \rho^\cut T^\cut(x) \big) \big(\pi T^\cut(x) \big)^4 u^\cut_{\mu}(x) u^\cut_{\nu}(x)  
                                      \\\nonumber&& \qquad  
                                      - \gamma^\cut \big(\rho^\cut T^\cut(x)\big) \big(\pi T^\cut(x)\big)^3 \sigma^\cut_{\mu\nu}(x) 
                                      + \mathcal{O}\left( \epsilon^2 \right) \Big)  \, .
\eea
The energy density $e^{\cut}$, the pressure $p^{\cut}$ and the shear viscosity $\eta^{\cut}$ at the cut-off are:
\be
e^{\cut} = \mathcal{K} \Big(\pi T^\cut \Big)^4 \left( \beta^{\cut} - \alpha^{\cut} \right) \, , \qquad  
p^{\cut} = \mathcal{K} \Big(\pi T^\cut \Big)^4 \alpha^{\cut}  \qquad \textrm{and} \qquad  
\eta^{\cut} = \mathcal{K} \Big(\pi T^\cut \Big)^3 \gamma^{\cut} \, .
\ee
Explicitly:
\begin{eqnarray}
\label{almostfinal}
e^\cut (\rho^\cut, x) &=& 3 \mathcal{K} \left( \pi T^\bd (x) \right)^4 \frac{ \partial_z A(z)\vert_{z=z^\cut}}{A(z^\cut)} 
    = \frac{3}{4} \mathcal{K}  \frac{\left( \pi T^\bd (x) \right)^4}{1+z^\cut/4}  \\\nonumber 
e^\cut (\rho^\cut, x) + p^\cut (\rho^\cut, x) &=& \mathcal{K} \left( \pi T^\bd (x) \right)^4 \frac{A(z^\cut)}{A(z^\cut)-B(z^\cut)} 
   \, \partial_z  \left. \left( \frac{ B(z)}{A(z)} \right) \right\vert_{z=z^\cut}
    = \mathcal{K} \frac{\left( \pi T^\bd (x) \right)^4}{(1-z^\cut/4)(1+z^\cut/4)}  \\\nonumber \quad 
\eta^\cut (\rho^\cut, x) &=&  \mathcal{K} \left( \pi T^\bd (x) \right)^3 \Big( A(z^\cut)-B(z^\cut) \Big)^{1/2}  
                            \partial_z \left. \left( \frac{ C(z)}{A(z)} \right) \right\vert_{z=z^\cut}
                          = \mathcal{K} \frac{\left( \pi T^\bd (x) \right)^3}{(1+z^\cut/4)^{3/2}} \cdot c_2 \, .
\end{eqnarray}
where $z$ and $z^\cut$ are defined as in \eqref{z}.
To derive these formulae we first plugged \eqref{metric}, \eqref{ucu} and \eqref{scs} into \eqref{TmunuCutoff}, then compared it with \eqref{tc} and \eqref{alphabetagamma} 
in order to express $e^\cut$, $p^\cut$ and $\eta^\cut$ in terms of $A(z)$, $B(z)$ and $C(z)$. Finally, we substituted the explicit solution of the equations of motion from \eqref{ABC} to arrive at the final result. It is important to emphasize here that $T^\bd(x)$ on the right hand side of all three equations in \eqref{almostfinal} should be seen as a function of $T^\cut(x)$ and $\rho^\cut$ (both in the overall factors and in the definition of $z^\cut$), 
since we want $e^\cut$, $p^\cut$ and $\eta^\cut$ to depend only on the cut-off
parameters $\rho^\cut$ and $T^\cut(x)$. The reason for this implicit dependence on $T^\cut(x)$ is 
the complicated inverse of the function given in \eqref{def2} (see the discussion below this equation).

It is worth to notice that both $e^\cut$ and $\eta^\cut$  are finite at the horizon ($z^\cut=4$), while the local pressure $p^\cut$ diverges (see \cite{Damour1, Damour2} for the earlier discussions). 

From \eqref{almostfinal} and the thermodynamic relation $e^\cut + p^\cut = T^\cut s^\cut$ we arrive at our final result:
\be
\frac{\eta^\cut}{s^\cut} = \frac{c_2}{\pi} \, .
\ee
Comparing with \eqref{etabd}, we learn that:
\be
\frac{\eta^\cut}{s^\cut} = \frac{\eta^\bd}{s^\bd} 
\ee
Thus, as advertised in Introduction, the physical ratio $\eta/s$ obtained after appropriate field redefinitions, does not flow radially irrespective of the explicit choice of 
data at $\rho_0$ that completely determines the radial flow (or equivalently the explicit choice of boundary conditions at the cut-off and at the horizon). 
Further, requiring regularity in the past/future horizon \eqref{hreg} implies that:
\be
\frac{\eta^\cut}{s^\cut} = \mp \frac{1}{4 \pi} \, .
\ee 
Thus the above is also true irrespective of the explicit boundary condition at the cut-off.  This result has been foreshadowed by the study of scale dependence of the holographic response function \cite{Iqbal, Nabamita}. It has also been derived in \cite{Strominger1} in the context of holographic non-relativistic hydrodynamics. \textit{Interestingly, again as in \cite{Strominger1} in the context of the holographic non-relativistic hydrodynamics,  the bulk viscosity also vanishes at all scales.} 

Additionally, we find that: 
\be 
\frac{s^\cut}{s^\bd} = \frac{1}{(1+z^\cut/4)^{3/2}} \, ,
\ee
therefore the entropy density decreases monotonically as move away from the boundary to the horizon. 
This is expected from the general renormalization group flow arguments.
The \emph{total} entropy and not the entropy density is specified by the area of the horizon. 
In the case of planar black holes the area of horizon is infinite, so the entropy density should 
scale as inverse of the $3d$ area. This is exactly what we find.

To  conclude, let us present the result for speed of sound at the cut-off, which follows directly from \eqref{almostfinal}:
\be
c^2_{\rm sound} = \frac{p^\cut}{e^\cut} = \frac{1}{3} \cdot  \frac{1 + 3 z^\cut/4}{1 - z^\cut/4}  \, .
\ee
We see that $c_{\rm sound}$ blows up at the horizon and is equal to $1/\sqrt{3}$ on the boundary in accordance with \cite{Damour1, Damour2, Eling1, Eling2, Strominger1}. The divergence of the speed of sound at the horizon ($z^\cut=4$) is most likely related to the incompressibility ($\nabla \cdot u = 0$) of the fluid on the horizon \cite{Eling2}.

\section{Discussion}
We have found that the renormalization group flow of energy-momentum tensor at a radial slice of space-time as given by Einstein's equations preserves the hydrodynamic form of the energy-momentum tensor up to first order in the derivative expansion for arbitrary slowly varying boundary data which equilibrate in the far past or far future. This result is $kinematic$, in the sense that to show this we only need to solve for the radial flow of the boundary data, without imposing the constraints implied by Einstein's equation on it. Also, this result is independent of the explicit choice of boundary conditions for the radial flow. Our results thus shed some light on holographic renormalization in general.

Much of our procedure can be generalized to higher derivative corrections to the Einstein-Hilbert action which are local functionals of the Riemann curvature and its covariant derivatives. However, we feel that we need to understand the renormalization group flow given by Einstein's equations better first. We mention a few open questions in this context:

\begin{enumerate}
\item 
It is important to figure out whether the hydrodynamic form of the renormalized energy-momentum tensor is preserved \textit{off-shell} by the radial flow of the boundary at second and higher orders in the derivative expansion as well. We expect this will help to address the general question of the choice of counter-terms and hypersurfaces involved in defining holographic renormalization.

\item
What are \emph{the degrees of freedom} at the horizon? In particular, we have to understand if the hydrodynamics at the horizon is purely incompressible Navier-Stokes. To address this problem, one needs to study the higher derivative corrections to the bulk metric. In principle, the answer can also depend on the renormalization scheme. 

\item
What is the renormalization group flow of the \emph{non-hydrodynamic} part of the boundary energy-momentum tensor? Non-hydrodynamic data at the boundary in the context of AdS/CFT has been investigated in \cite{myself2, myself3}.
The study of the renormalization group flow of such non-hydrodynamic data will help to answer the question if non-hydrodynamic degrees of freedom are relevant at a regular future horizon.

\item
The study of RG flow of the energy-momentum tensor in the hydrodynamic regime itself may throw light on \emph{the holographic emergence of gravity} as argued in \cite{Jacobson, Padmanabhan1, Verlinde, Padmanabhan2}. In particular, we know if that if the transport coefficients of the holographic fluid are not right, we end up with naked singularities. If naked singularities are unobservable, we need to understand this in the context of holographic emergence of gravity. It may be possible that absence of naked singularities comes from appropriate physical restriction on the renormalization group flow of the boundary data.    
\end{enumerate}

\section*{Acknowledgments}
It is a pleasure to thank Giuseppe Policastro and Miguel Paulos for useful and enjoyable discussions during the course of this work.
The works of S.~K. is supported in part by the European Commission Marie Curie Fellowship under the
contracts IEF-2008-237488. The research of AM is supported by the grant number ANR-07-CEXC-006 of L'Agence Nationale de La Recherche.

\appendix
\section{Translation to Eddington-Finkelstein coordinates}

The metric can be translated to the ingoing/outgoing Eddington-Finkelstein coordinates order by order in the derivative expansion. The coordinate transformations also receive corrections in the derivative expansion which can be systematically computed. We will denote the Eddington-Finkelstein radial coordinate as $r$ and the boundary coordinates of this coordinate system collectively as $y$. Following \cite{myself1}, we find that up to first order in the derivative expansion, in the ingoing Eddington-Finkelstein coordinates the metric \eqref{metric} is:
\begin{eqnarray}\label{efm}
ds^2 &=& -2u_\mu(y) dy^\mu dr + G_{\mu\nu}(r, y) dy^\mu dy^\nu, \\\nonumber
G_{\mu\nu}(y,r) &=& r^2 h_{\mu\nu}(y) + \frac{(\pi T(y))^4}{r^2}u_\mu (y) u_\nu (y) 
\\\nonumber&& + \frac{r^2}{2\pi T(y)}\Bigg(\frac{\left(r+\pi T(y)\right)^2\left(r^2 + (\pi T(y))^2\right)}{r^4} -2\arctan \left(\frac{r}{\pi T(y)}\right) +\pi \Bigg) \sigma_{\mu\nu}(y) 
\\\nonumber&& - r\Bigg((u(y)\cdot\nabla)(u_\mu(y)u_\nu(y)) -\frac{2}{3}u_\mu(y)u_\nu(y)(\nabla\cdot u)(y)\Bigg)
\\\nonumber&& + \tilde{c}_1 \Bigg(\frac{T(y)}{r_0}\Bigg)r^2 \sigma_{\mu\nu}(y) + \Bigg(\tilde{c}_2 \Bigg(\frac {T(y)}{r_0}\Bigg)-\frac{1}{4}\Bigg) \frac{r^2}{\pi T(y)} \log \Bigg(1- \frac{(\pi T(y))^4}{r^4}\Bigg)\sigma_{\mu\nu}(y) 
\\\nonumber&&+ \mathcal{O}\left( \epsilon^2 \right) \, ,
\end{eqnarray} 
where:
\be
\tilde{c}_1 \Bigg(\frac{T(y)}{r_0}\Bigg) = c_1 (\rho_0 T(x)) + \mathcal{O}\left( \epsilon \right), \quad \tilde{c}_2 \Bigg(\frac{T(y)}{r_0}\Bigg) = c_2 (\rho_0 T(x)) + \mathcal{O}\left( \epsilon \right),
\ee
parametrizing the choice of data at the Eddington-Finkelstein radial slice $r_0$ just as at the Fefferman-Graham radial slice $\rho_0$ in Section 2.

In the above coordinates the horizon is at $r=\pi T(y)$, where $G_{\mu\nu}u^\mu(y) u^\nu(y)$ vanishes. Also the boundary data ($u^\mu(y), T(y), h_{\mu\nu}(y)$) are the same functions of new boundary coordinates $y$. The boundary is located at $r=\infty$ where the new boundary coordinates $y$ coincide with the Fefferman-Graham boundary coordinates $x$, though they are nontrivially related in the bulk. Except for the penultimate line in \eqref{efm}, this metric agrees with that in \cite{Bhattacharyya4}, where arbitrary perturbations of the induced metric and the Brown-York tensor have not been considered as here. 

We also see from the explicit form of the metric in the ingoing Eddington-Finkelstein coordinates that the future horizon is regular when $c_2 = 1/4$, \emph{i.e.} when $\eta^b /s^b = 1/(4\pi)$. The metric in outgoing Eddington-Finkelstein coordinates can be obtained by reversing the sign of the time coordinate and also the time component of $u^\mu (y)$. We readily observe that for regularity of the past horizon, we need $\eta^b/s^b = - 1/(4\pi)$. The reversal of the sign simply follows from the time reversal invariance of Einstein's equations.


\end{document}